# THE BARISTA: A MODEL FOR BID ARRIVALS IN ONLINE AUCTIONS

By Galit Shmueli, Ralph P. Russo and Wolfgang Jank

*University of Maryland, University of Iowa and University of Maryland*

The arrival process of bidders and bids in online auctions is important for studying and modeling supply and demand in the online marketplace. A popular assumption in the online auction literature is that a Poisson bidder arrival process is a reasonable approximation. This approximation underlies theoretical derivations, statistical models and simulations used in field studies. However, when it comes to the bid arrivals, empirical research has shown that the process is far from Poisson, with early bidding and last-moment bids taking place. An additional feature that has been reported by various authors is an apparent self-similarity in the bid arrival process. Despite the wide evidence for the changing bidding intensities and the self-similarity, there has been no rigorous attempt at developing a model that adequately approximates bid arrivals and accounts for these features. The goal of this paper is to introduce a family of distributions that well-approximate the bid time distribution in hard-close auctions. We call this the BARISTA process (Bid ARrivals In STAges) because of its ability to generate different intensities at different stages. We describe the properties of this model, show how to simulate bid arrivals from it, and how to use it for estimation and inference. We illustrate its power and usefulness by fitting simulated and real data from eBay.com. Finally, we show how a Poisson bidder arrival process relates to a BARISTA bid arrival process.

**1. Introduction and motivation.** Empirical research of online auctions has been flourishing in recent years due to the important role that these auctions play in the marketplace, and the availability of large amounts of high-quality bid data from websites such as eBay, Yahoo!, OnSale and uBid. Many of the theoretical results derived for traditional (offline) auctions have been shown to fail in the online setting for reasons such as globalism, computerized bidding and the recording of complex bids, longer auction durations, more flexibility in design choice by the seller and issues of trust. A central factor underlying many important results is the number of bidders participating in the auction. Typically, it is assumed that this number is fixed [Pinker et al. (2003)] or fixed but unknown [McAfee and McMillan (1997)]. In online auctions the number of bidders and bids is not predetermined, and it is known to be affected by the auction design and its dynamics. Thus, in both the theoretical and empirical domains the number of bidders and bids plays an important role.

We propose a new and flexible model for the bid arrival process. Having a model for bid arrivals has several important implications. First, many researchers in the online auction arena use simulated bid arrival data to validate their results. Bapna et al. (2002), for example, use simulated bid arrival data to validate their model on a bidder's willingness to pay. Gwebu et al. (2005) design a complex simulation study to analyze bidders' strategies using assumptions about the bidder, as well as bid arrival rates. It has also been noted that the placement of bids influences the bidder arrival process [Beam et al. (1996)]. Hlasny (2006) reviews several econometric procedures for testing for the presence of latent shill-bidding (where sellers fraudulently bid on their own item) based on the arrival rate of bids. While a clear understanding of the process of bidding can have an impact on the theoretical literature, it can also be useful in many applications. These range from automated electronic negotiation through monitoring auction server performance to designing automated bidding agents. For instance, Menasce and Akula (2004) study the connection between bid arrivals and auction server performance. They find that the commonly encountered "last minute bidding" creates a huge surge in the workload of auction servers and degrades their performance. They then suggest a model to improve a server's performance through auction rescheduling using simulated bid arrival data.

Modeling the bid arrival process rather than the bidder arrivals also promises to produce more reliable results, because bid placements are typically completely observable from the auction's bid history, whereas bidder arrivals are not. eBay, for instance, posts the temporal sequence of all the bids placed over the course of the auction. In particular, every time a bid is placed its exact time-stamp is posted. In contrast, the time when bidders first arrive at an auction is unobservable from the bid history. Bidders can browse an auction without placing a bid, thereby not leaving a trace or revealing their interest in that auction. That is, they can look at a particular auction, inform themselves about current bid- and competition-levels in that auction, and make decisions about their bidding strategies. All this activity can take place without leaving an observable trace in the bid history that the auction site makes public. In fact, it is likely that bidders first browse an auction and only later place their bid. The gap between the bidder arrival time and bid placement also means that the bidder arrival is not identical to the bid arrival, and can therefore not be inferred directly from the observed bid times. Another issue is that most online auctions allow bid-revision, and therefore many bidders place multiple bids. This further adds to the obscurity of defining the bidder arrival-departure process. Our approach is therefore to model the bid arrival process based on empirical evidence.

The current literature, based on publicly-available bid data, reports strong evidence of two major features of the bid arrival process in online auctions: (1) a nonhomogenous intensity that possesses two or three distinct stages, and (2) a self-similarity effect in the distribution of bid arrival times. We describe these in Section 2. However, aside from noting these features, no model has been suggested for approximating the bid arrival process that addresses these two features. In light of the absence of such a model, we introduce the BARISTA process, a model that well-approximates bid arrivals in online auctions. Section 3 introduces the model and its properties, and describes two special cases. Section 4 describes a method for simulating data from this process and several methods for estimating model parameters. In Section 5 we use simulated data and a diverse set of real bid data from eBay to illustrate the estimation and model fit. In addition to the various uses of the bid arrival model, one might be able to infer bidder strategies from the aggregate bid arrival process. In Section 6 we tie the bid and bidder arrival processes, proposing several bidding strategies that would lead to BARISTA-type bid arrivals. In Section 7 we conclude and suggest future enhancements.

**2. Features of bid arrivals.** We start by describing two prominent features of bid arrivals that have been reported in the literature, and follow with an illustration using bid data from eBay.

2.1. *Multi-stage arrival intensities.* Time-limited tasks are omnipresent in the offline world: voting for a new president, purchasing tickets for a popular movie or sporting event, filing one's federal taxes, etc. In many of these cases arrivals are especially intense as the deadline approaches. For instance, during the 2001 political elections in Italy, more than 20 million voters cast their ballots between 13:00–22:00 [Bruschi et al. (2002)], when ballots were scheduled to close at 22:00. Similarly, a high proportion of U.S. tax returns are filed near the 15 April deadline. For instance, about one-third of all returns are not filed until the last two weeks of tax season (www.heraldstandard.com/site/news.cfm?newsid=14359378&BRD=2280&PAG=461&dept_id=480247 According to Ariely et al. (2005), deadline effects have been noted in studies of bargaining, where agreements are reached in the final moments before the deadline [Roth et al. (1998)]. Such effects have been shown among animals, which respond more vigorously toward the expected end of a reinforcement schedule, and in human task completion where individuals become increasingly impatient toward the task's end. Furthermore, people use different strategies when games are framed as getting close to the end
[even when these are arbitrary break points; Croson (1996)]. In addition to
the deadline effect, there is an effect of earliness where the strategic use of
time moves transactions earlier than later, for example, in the labor market
[Roth and Xing (1994); Avery et al. (2001)].

Such deadline and earliness effects have also been observed in the online environment. Several researchers have noted deadline effects in internet
auctions [Bajari and Hortacsu (2000); Borle et al. (2006); Ku et al. (2004);
Roth and Ockenfels (2000); Wilcox (2000)]. In many of these studies it was
observed that a nonnegligible percent of bids arrive at the very last minute
of the auction. This phenomenon, called "bid sniping," has received much
attention, and numerous explanations have been suggested to explain its
existence. Empirical studies of online auctions have also reported an unusual amount of bidding activity at the auction start followed by a longer
period of little or no activity [Borle et al. (2006); Jank and Shmueli (2007)].
Bapna et al. (2003) refer to bidders who place a single early bid as "evaluators." Finally, "bid shilling," a fraudulent act where the seller places
dummy bids to drive up the price, is associated with early and high bidding
[Kauffman and Wood (2000)]. The existence of these bid-timing phenomena
are important factors in determining outcomes at the auction level, as well
as at the market level. They have therefore received much attention from
the research community.

2.2. *Self-similarity (and its breakdown).* While both the offline and online environments share the deadline and earliness effects, the online environment appears to possess the additional property of *self-similarity* in the bid
arrival process [this property was also found in the offline process of bargaining agreements, as described in Roth and Ockenfels (2000)]. Self-similarity
refers to the "striking regularity" of shape that can be found among the distribution of bid arrivals over the intervals $[t, T]$, as $t$ approaches the auction
deadline $T$. Self-similarity is central in applications such as web, network and
ethernet traffic. Huberman and Adamic (1999) found that the number of
visitors to websites follows a universal power law. Liebovitch and Schwartz
(2003) reported that the arrival process of email viruses is self-similar. However, this has also been reported in other online environments. For instance,
Aurell and Hemmingsson (1997) showed that times between bids in the interbank foreign exchange market follow a power law distribution.

Several authors reported results that indicate the presence of self-similarity
in the bidding frequency in online auctions. Roth and Ockenfels (2000) found
that the arrival of last bids by bidders during an online auction is closely
related to a self-similar process. They approximated the CDF of bid arrivals
in "reverse time" (i.e., the CDF of the elapsed times between the bid arrivals and the auction deadline) by the power functional form $F_T(t) = (t/T)^\alpha$ ($\alpha > 0$), over the interval $[0,T]$, and estimated $\alpha$ from the data using OLS. This approximates the distribution of bids over intervals that range from the last 12 hours to the last 10 minutes, but accounts for neither the final minutes of the auction nor the auction start and middle. Yang et al. (2003) found that the number of bids and the number of bidders in auctions on eBay and on its Korean partner (auction.co.kr) follow a power law distribution. This was found for auctions across multiple categories. The importance of this finding, which is closely related to the self-similarity property, is that the more bidding one observes up to a fixed time point, the higher the likelihood of seeing another bid before the next time point. According to Yang et al. (2003), such power-law behaviors imply that the online auction system is driven by self-organized processes, involving all bidders who participate in a given auction activity.

The implications of bid arrivals following a self-similar process instead of an ordinary Poisson model are significant: The levels of activity throughout an auction with self-similar bid arrivals would increase at a much faster rate than expected under a Poisson model. It would be especially meaningful toward the end of the auction, which has a large impact on the bid amount process and the final price. The self-similar property suggests that the rate of incoming bids increases steadily as the auction approaches its end. Indeed, empirical investigations have found that many bidders wait until the very last possible moment to submit their final bid. By doing so, they hope to increase their chance of winning the auction since the probability that another competitor successfully places an even higher bid before closing is diminishing. This common bidding strategy of "bid sniping" (or "last minute bidding") would suggest a steadily increasing flow of bid arrivals toward the auction end. However, empirical evidence from online auction data indicates that bid times over the last minute or so of hard-close auctions tend to follow a uniform distribution [Roth et al. (1998)]. This has not been found in soft-close, or "going-going-gone" auctions, such as those on Amazon, Yahoo! or uBid.com, where the auction continues several minutes after the last bid is placed.

Thus, in addition to the evidence for self-similarity in online auctions, there is also evidence of its breakdown during the very last moments of a hard-close auction. Roth and Ockenfels (2000) note that the empirical CDF plots for intervals that range between the last 12 hours of the auction and the last 1 minute all look very similar except for the last 1-minute plot. Being able to model this breakdown is essential, since the last moments of the auction (when sniping takes place) are known to be crucial to the final price. In the absence of such a model, we introduce a bid arrival process that describes the frequency throughout the *entire* auction. Rather than focusing on the last several hours and excluding the last moments, our model accommodates the changes in bidding intensity from the very start to the very end of a hard-close auction.

To illustrate the self-similarity in the bid arrival process in online auctions, we collected data on 189 7-day auctions (with a total of 3651 bid times) on eBay.com for new Palm M515 Personal Digital Assistants. Figure 1 displays the empirical CDFs for the 3651 bid arrivals for the purposes of examining the self-similarity property. The CDF is plotted at several resolutions, "zooming-in" from the entire auction duration (of 7 days) to the last day, last 12-hours, 6-hours, 3-hours, 5-minutes, 2-minutes and the very last minute. We see that (1) the last day curve (thick black) is different from the other curves in that it starts out concave, (2) the last day through last 3-hour curves (red) are all very similar to each other, and (3) the last minutes curves (grey) gradually approach the 1-minute curve which is nearly uniform. These visual similarities are confirmed by the results of two-sample Kolmogorov–Smirnoff tests where we compared all the pairs of distributions and found similarities only among the curves within each of the three groups.

This replicates the results in Roth and Ockenfels (2000) where self-similarity was observed in the bid time distributions of the last 12-hour, 6-hour, 3-hour, 1-hour, 30-minute and 10-minute periods of the auction, and where this self-similarity breaks down in the last minute of the auction to become a uniform distribution. However, we examine a few additional time resolutions which give further insight: First, by looking at the last 5-minutes and last 2-minutes bid distributions, we see that the self-similarity gradually transitions into the 1-minute uniform distribution. Second, our inspection of the entire auction duration [which was unavailable in the study by Roth and Ockenfels (2000)] reveals an additional early-bidding stage. Self-similarity, it appears, is not prevalent throughout the entire auction duration! Such a phenomenon can occur if the probability of a bid not getting registered on the auction site is positive at the last moments of the auction, and increases as the auction comes to a close. There are various factors that may cause a bid to not get registered. One possible reason is the time it takes to manually place a bid [Roth and Ockenfels (2000) found that most last minute bidders tend to place their bids manually rather than through available sniping software agents]. Other reasons are hardware difficulties, Internet congestion, unexpected latency and server problems on eBay (see, e.g., www.auctionsniper.com). Clearly, the closer to the end the auction gets, the higher the likelihood that a bid will not get registered successfully. This increasing likelihood of an unsuccessful bid counteracts the increasing flow of last minute bids. The result is a uniform bid arrival process that "contaminates" the self-similarity, of the arrivals until that point.

In the next section we introduce a flexible nonhomogeneous Poisson process (NHPP) that captures the empirical phenomenon described above. In addition to the self-similarity, it also accounts for the two observed phenomena of "early bidding" and "last minute bidding" (sniping).

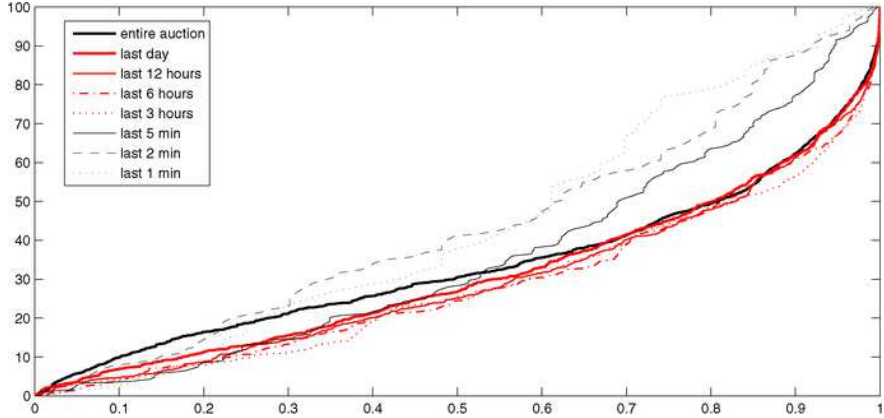

Fig. 1. *Empirical CDFs of number of bids in 189 Palm M515 auctions overlaid.*

**3. The BARISTA: A three-stage nonhomogeneous Poisson process.** We introduce a process that captures the two main features of arrivals in online auctions: the three stages and the self-similarity (with its breakdown). We call this the BARISTA (Bid ARrivals In STAges) process, because it generates different intensities of activity [we also call the stages the "espresso stage" (short and intense), "macchiato stage" (stained) and "ristretto stage" (extra intense), and hence the BARISTA]:

*Stage* 1. The auction start, characterized by an early burst of activity,

*Stage* 2. The mid-auction bid arrivals, characterized by increasing bid intensity and self-similarity that is gradually contaminated as the 3rd stage is approached, and

*Stage* 3. The last moments of the auction, characterized by very intense activity dampened possibly by bids that are not successfully transmitted.

3.1. *Model formulation.* A nonhomogeneous Poisson process differs from an ordinary Poisson process in that its intensity is not a constant, but rather a function of time. We introduce a particular intensity function that captures the three-stage dynamics described above. Suppose bids arrive during $[0, T]$ in accordance with a nonhomogeneous Poisson process $N(s)$, $0 \leq s \leq T$, with intensity function

$$(1) \quad \lambda(s) = \begin{cases} c\left(1 - \dfrac{d_1}{T}\right)^{\alpha_2 - \alpha_1} \left(1 - \dfrac{s}{T}\right)^{\alpha_1 - 1}, & \text{for } 0 \leq s \leq d_1, \\ c\left(1 - \dfrac{s}{T}\right)^{\alpha_2 - 1}, & \text{for } d_1 \leq s \leq T - d_2, \\ c\left(\dfrac{d_2}{T}\right)^{\alpha_2 - \alpha_3} \left(1 - \dfrac{s}{T}\right)^{\alpha_3 - 1}, & \text{for } T - d_2 \leq s \leq T, \end{cases}$$

where $c > 0$, $\alpha_j > 0$ for $j = 1, 2, 3$, $T$ is a known constant, and $0 \leq d_1 < T - d_2 \leq T$. Note that this intensity function is continuous, so there are no jumps at times $d_1$ and $T - d_2$. The random variable $N(s)$ which counts the number of arrivals until time $s$ follows a Poisson distribution with mean

$$
(2) \quad m(s) = \begin{cases} K\left(1 - \left(1 - \dfrac{s}{T}\right)^{\alpha_1}\right), & \text{for } 0 \leq s \leq d_1 \\ K\left(1 - \left(1 - \dfrac{d_1}{T}\right)^{\alpha_1}\right) + \dfrac{Tc}{\alpha_2}\left(\left(1 - \dfrac{d_1}{T}\right)^{\alpha_2} - \left(1 - \dfrac{s}{T}\right)^{\alpha_2}\right), \\ \qquad\qquad\qquad\qquad\qquad\text{for } d_1 \leq s \leq T - d_2, \\ K\left(1 - \left(1 - \dfrac{d_1}{T}\right)^{\alpha_1}\right) + \dfrac{Tc}{\alpha_2}\left(\left(1 - \dfrac{d_1}{T}\right)^{\alpha_2} - \left(\dfrac{d_2}{T}\right)^{\alpha_2}\right) \\ \quad + \dfrac{Tc}{\alpha_3}\left(\dfrac{d_2}{T}\right)^{\alpha_2 - \alpha_3}\left(\left(\dfrac{d_2}{T}\right)^{\alpha_3} - \left(1 - \dfrac{s}{T}\right)^{\alpha_3}\right), \\ \qquad\qquad\qquad\qquad\qquad\text{for } T - d_2 \leq s \leq T, \end{cases}
$$

where $K = \dfrac{Tc}{\alpha_1}(1 - \dfrac{d_1}{T})^{\alpha_2 - \alpha_1}$.

Given that $N(T) = n$, the collection of arrival times are equivalent to the order statistics of a random sample of size $n$ from the distribution having distribution $F(s) = m(s)/m(T)$:

$$
(3) \quad F(s) = \begin{cases} \dfrac{CT}{\alpha_1}\left(1 - \dfrac{d_1}{T}\right)^{\alpha_2 - \alpha_1}\left[1 - \left(1 - \dfrac{s}{T}\right)^{\alpha_1}\right], \\ \qquad\qquad\qquad\qquad\qquad\text{for } 0 \leq s \leq d_1, \\ \dfrac{CT}{\alpha_1\alpha_2}\bigg[(\alpha_1 - \alpha_2)\left(1 - \dfrac{d_1}{T}\right)^{\alpha_2} \\ \quad + \alpha_2\left(1 - \dfrac{d_1}{T}\right)^{\alpha_2 - \alpha_1} - \alpha_1\left(1 - \dfrac{s}{T}\right)^{\alpha_2}\bigg], \\ \qquad\qquad\qquad\qquad\qquad\text{for } d_1 \leq s \leq T - d_2, \\ 1 - \dfrac{CT}{\alpha_3}\left(\dfrac{d_2}{T}\right)^{\alpha_2 - \alpha_3}\left(1 - \dfrac{s}{T}\right)^{\alpha_3}, \qquad \text{for } T - d_2 \leq s \leq T. \end{cases}
$$

Note that for the interval $d_1 \leq s \leq T - d_2$ we can write the CDF as

$$
(4) \qquad F(s) = F(d_1) + \dfrac{CT}{\alpha_2}\left[\left(1 - \dfrac{d_1}{T}\right)^{\alpha_2} - \left(1 - \dfrac{s}{T}\right)^{\alpha_2}\right],
$$

where

$$
C = c/m(T)
$$
$$
= \dfrac{\alpha_1\alpha_2\alpha_3/T}{(1 - d_1/T)^{\alpha_2}\alpha_3(\alpha_1 - \alpha_2) + \alpha_3\alpha_2(1 - d_1/T)^{\alpha_2 - \alpha_1} + (d_2/T)^{\alpha_2}\alpha_1(\alpha_2 - \alpha_3)}.
$$

The density function corresponding to this process is given by

$$(5) \quad f(s) = \begin{cases} C\left(1-\dfrac{d_1}{T}\right)^{\alpha_2-\alpha_1}\left(1-\dfrac{s}{T}\right)^{\alpha_1-1}, & \text{for } 0 \leq s \leq d_1, \\ C\left(1-\dfrac{s}{T}\right)^{\alpha_2-1}, & \text{for } d_1 \leq s \leq T-d_2, \\ C\left(\dfrac{d_2}{T}\right)^{\alpha_2-\alpha_3}\left(1-\dfrac{s}{T}\right)^{\alpha_3-1}, & \text{for } T-d_2 \leq s \leq T. \end{cases}$$

We expect $\alpha_3$ to be close to 1 (uniform arrival of bids at the end of the auction) and $\alpha_1 > 1$ to represent the early surge in bidding.

3.2. *Properties of the BARISTA process.* The process described by (1)–(4) has two properties that lead to a wide family of processes, and that can be useful in practice. We describe each property and its implications below.

3.2.1. *An additive property.* If $N_k$, $1 \leq k \leq m$, are independent BARISTA processes having $c$ parameters $c_1, \ldots, c_m$ and common $(\alpha_1, \alpha_2, \alpha_3)$, $(d_1, d_2)$ and $T$ parameters, then the aggregated process $N = \sum_{1 \leq k \leq m} N_k$ is a BARISTA with parameters $(\alpha_1, \alpha_2, \alpha_3)$, $(d_1, d_2)$, $T$, and $c = \sum_{1 \leq k \leq m} c_k$.

This means that the bid arrival times from several auctions may be aggregated and treated as though they were generated by a single auction, provided that each original auction can be regarded as producing a BARISTA process with the same (or nearly the same) parameters. The advantage of aggregation is more accurate parameter estimation.

3.2.2. *A regenerative property.* An observer who counts only the bid arrivals occurring after time $\beta T$, some $0 \leq \beta < 1$, sees the process

$$N_\beta(s) := N(s) - N(\beta T), \qquad \beta T \leq s \leq T.$$

$N_\beta$ is an *NHPP* with intensity function $\lambda_\beta = \lambda$, restricted to the interval $[\beta T, T]$.

Taking $\beta T$ as the new *zero*, and recording time on a new (faster) clock where *one new minute (a shminute) = $(1-\beta)$ minutes* on the original clock, we can write $\lambda_\beta$ as

$$\lambda_\beta(s) = \begin{cases} c_\beta\left(1-\dfrac{d_{1,\beta}}{T_\beta}\right)^{\alpha_2-\alpha_1}\left(1-\dfrac{s}{T}\right)^{\alpha_1-1}, & 0 \leq s \leq d_{1,\beta}, \\ c_\beta\left(1-\dfrac{s}{T_\beta}\right)^{\alpha_2-1}, & d_{1,\beta} \leq s \leq T_\beta - d_{2,\beta}, \\ c_\beta\left(\dfrac{d_{2,\beta}}{T_\beta}\right)^{\alpha_2-\alpha_3}\left(1-\dfrac{s}{T_\beta}\right)^{\alpha_3-1}, & T_\beta - d_{2,\beta} \leq s \leq T_\beta, \end{cases}$$

where $c_\beta = c(1-\beta)^{\alpha_2-1}$, $d_{1,\beta} = \max(d_1 - \beta T, 0)$, $d_{2,\beta} = \min(d_2, T_\beta)$, and $T_\beta = (1-\beta)T$. Thus, $\lambda_\beta$ has the same form as our original $\lambda$ with $\lambda = \lambda_0$ in the new notation.

This regenerative property means that we can use the BARISTA model to approximate bid arrivals in an ongoing auction, not only in a completed auction. One application where this is useful is in real-time forecasting of future bid times, such as for the purpose of optimizing server performance.

3.3. *Special cases.* In empirical studies, $d_1$ appears to be small (1–2 days) and $d_2$ very small (a few minutes) compared to $T$ (several days). Thus, most of the BARISTA process is realized in the second stage, during which the process can be regarded as having *contaminated self-similarity*. The contamination is caused by the bid arrivals in the third stage, and increases as $s \to T - d_2$.

When $d_1 = d_2 = 0$, the BARISTA process reduces to a single-stage process (NHPP$_1$) with an intensity function $\lambda(s) = c(1 - \frac{s}{T})^{\alpha-1}$ and associated *CDF* function $F(s) = 1 - (1 - \frac{s}{T})^\alpha, 0 \leq s \leq T$. For $(\theta, t) \in [0,1] \times (0,T]$, we have

$$\frac{1 - F(T - t\theta)}{1 - F(T - t)} = \theta^\alpha \qquad \text{(independent of } t),$$

and thus, we have a *pure self-similar* process. The joint *MLE* of $(\alpha, c)$ is obtainable in this case (see Appendix A):

$$\widehat{\alpha} = -N(T)\left[\sum_{i=1}^{N(T)} \ln\left(1 - \frac{X_i}{T}\right)\right]^{-1}, \qquad \widehat{c} = \frac{N(T)\widehat{\alpha}}{T}.$$

Since $X \sim F \Longrightarrow -\ln(1 - \frac{X}{T}) \sim \exp(rate = \alpha)$, and $\lim_{c \to \infty} \Pr(N(T) \to \infty) = 1$, a conditioning argument on $N(T)$ yields an asymptotic result:

$$\sqrt{N(T)}\left(\frac{\alpha}{\widehat{\alpha}} - 1\right) \xrightarrow{D} N(0,1) \qquad \text{as } c \to \infty,$$

where $N(0,1)$ indicates a standard normal distribution.

When $d_1 = 0$, the BARISTA process reduces to a two-stage process (NHPP$_2$), with a single *changepoint* at $T - d_2$. This process is useful for modeling bid arrivals in auctions that lack the initial surge of early bidding. For further technical details on these special cases, see Shmueli et al. (2004).

**4. Fitting the BARISTA to data.** Simulated bid arrivals are useful in field experiments, in evaluation of model fit, and for quantifying sampling error. The method is simple to program and computationally efficient. Fitting the BARISTA process to data requires estimating the two changepoints and three $\alpha$ parameters. We introduce three estimation methods that range in their computational intensiveness and accuracy (Matlab code for the simulation and estimation procedures is available at http://www.smith.umd.edu/ceme/statistics/code.html).

4.1. *Process simulation.* To simulate $n$ observations from the BARISTA process on the interval $[0, T]$, we use the inversion method and apply the inverse CDF to a simulated random sample of U(0, 1) variates. In particular, the inverse CDF can be written as

$$
(6) \quad F^{-1}(s) = \begin{cases} T - T\left\{1 - \dfrac{s\alpha_1}{CT}\left(1 - \dfrac{d_1}{T}\right)^{\alpha_1 - \alpha_2}\right\}^{1/\alpha_1}, \\ \qquad\qquad\qquad\qquad\qquad\qquad \text{for } 0 \leq s \leq d_1, \\ T - T\left\{\left(1 - \dfrac{d_1}{T}\right)^{\alpha_2} - \dfrac{\alpha_2}{CT}(s - F_3(d_1))\right\}^{1/\alpha_2}, \\ \qquad\qquad\qquad\qquad\qquad\qquad \text{for } d_1 \leq s \leq T - d_2, \\ T - T\left\{\dfrac{\alpha_3}{CT}(1 - s)\left(\dfrac{d_2}{T}\right)^{\alpha_3 - \alpha_2}\right\}^{1/\alpha_3}, \qquad \text{for } T - d_2 \leq s \leq T. \end{cases}
$$

The algorithm for generating $n$ arrivals $(x_1, \ldots, x_n)$ is then:
(1) Generate $n$ uniform variates $u_1, \ldots, u_n$.
(2) For $k = 1, \ldots, n$, set

$$
(7) \quad x_k = \begin{cases} T - T\left\{1 - \dfrac{u_k\alpha_1}{CT}\left(1 - \dfrac{d_1}{T}\right)^{\alpha_1 - \alpha_2}\right\}^{1/\alpha_1}, \\ \qquad\qquad\qquad\qquad\qquad\qquad \text{if } u_k < F(d_1), \\ T - T\left\{\dfrac{\alpha_2}{CT}(F_3(d_1) - u_k) + \left(1 - \dfrac{d_1}{T}\right)^{\alpha_2}\right\}^{1/\alpha_2}, \\ \qquad\qquad\qquad\qquad\qquad\qquad \text{if } F(d_1) \leq u_k < F(T - d_2), \\ T - T\left\{\dfrac{\alpha_3}{CT}u_k\left(\dfrac{d_2}{T}\right)^{\alpha_3 - \alpha_2}\right\}^{1/\alpha_3}, \qquad \text{if } u_k \geq F(T - d_2). \end{cases}
$$

Note that we fix the number of bid arrivals ($n$) rather than a randomly generated number, since the estimators are of the same form in both cases (see Appendix A). In order to generate a random number of bids under the BARISTA model, one would generate a variate from a Poisson distribution with mean $m(T)$ [see equation (2)].

4.2. *Parameter estimation.* We describe three estimation methods each having a different tradeoff between computational intensity and accuracy, and with varying amounts of required user input.

4.2.1. *Quick and crude (CDF-based) estimation.* The estimation of the $\alpha$ parameters depends on the changepoints $d_1$, $T - d_2$ and vice-versa. As a crude start, we choose three intervals of the form $[T - t, T - s]$ that we are confident lie in the first, second or third stages, and use those for estimating the $\alpha$ parameters. We then use the $\alpha$ estimates to obtain estimates for the changepoints.

In both cases the estimates are based on writing the parameters as a function of the CDF, and then substituting the empirical CDF to obtain estimates.

*Estimation of $\alpha$ parameters.* From (3) it can be seen that in each interval the CDF of the BARISTA process is in the form $F(t) = \beta_j - \theta_j(1 - \frac{t}{T})^{\alpha_j}$, $(j = 1, 2, 3)$, and therefore the same approximation works on each of the three intervals $[0, d_1]$, $[d_1, T - d_2]$ and $[T - d_2, T]$. After choosing intervals $[T - t, T - s]$ that we are confident lie in stage $j$ (the first, second or third stage), we have

$$
\begin{aligned}
(8) \quad & \frac{F(T-t) - F(T - \sqrt{st})}{F(T - \sqrt{st}) - F(T-s)} \\
&= \frac{\theta_j[1 - (1 - (T-t)/T)^{\alpha_j}] - \theta_j[1 - (1 - (T-\sqrt{st})/T)^{\alpha_j}]}{\theta_j[1 - (1 - (T-\sqrt{st})/T)^{\alpha_j}] - \theta_j[1 - (1 - (T-s)/T)^{\alpha_j}]} \\
&= \frac{(ts)^{\alpha_j/2} - t^{\alpha_j}}{s^{\alpha_j} - (st)^{\alpha_j/2}} = \frac{(s^{\alpha_j/2} - t^{\alpha_j/2})t^{\alpha_j/2}}{(s^{\alpha_j/2} - t^{\alpha_j/2})s^{\alpha_j/2}} = \left(\frac{t}{s}\right)^{\alpha_j/2}.
\end{aligned}
$$

The relevant $\alpha$ is given by

$$
(9) \quad \alpha_j = 2 \frac{\ln[F(T-t) - F(T-\sqrt{st})] - \ln[F(T-\sqrt{st}) - F(T-s)]}{\ln t - \ln s}.
$$

We then estimate $\alpha_j$ by substituting $F$ with the empirical CDF $F_e = N(t)/N(T)$ in the approximation.

For $\alpha_3$, we can use the exact relation

$$
(10) \quad \alpha_3 = \frac{\ln[R(t_3)/R(t_3')]}{\ln[(T-t_3)/(T-t_3')]},
$$

where $R(t) = 1 - F(t)$ and $t_3, t_3'$ are within $[T - d_2, T]$. To estimate $\alpha_3$, we choose reasonable values of $t_3, t_3'$ and use the empirical survival function $R_e = 1 - F_e$.

Obtaining standard errors for these estimators can be done by bootstrapping [see Efron and Tibshirani (1993) for details], due to the low computational effort involved in this estimation method.

To assess this method, we simulated 5000 random observations from the BARISTA process on the interval $[0, 7]$ with parameters $\alpha_1 = 3, \alpha_2 = 0.4, \alpha_3 = 1$ and the changepoints $d_1 = 2.5$ (defining the first 2.5 days as the first stage) and $d_2 = 5/10080$ (defining the last 5 minutes as the third stage). The intensity function for these data is shown in Figure 2, and parameter estimates with their standard errors are given in Table 1.

To study the robustness of the estimators to the choice of $t$ and $s$, we computed the quick and crude estimate for $\alpha_1$ on a range of intervals of the form $[0.001, t_1]$, where $0.5 \leq t_1 \leq 5$. Note that this interval includes values that are outside the range $[0, d_1 = 2.5]$. The left panel in Figure 3 illustrates the estimates obtained for these intervals. For values of $t_1$ between 1.5–3.5 days, the estimate for $\alpha_1$ is relatively stable and close to 3. Similarly, the right panel in Figure 3 describes the estimates of $\alpha_3$, using (10), as a function of the choice of $t_3$ with $t'_3 = 7 - 1/10080$. The estimate is relatively stable and close to 1.

For estimating $\alpha_2$, an interval such as $[3, 6.9]$ is reasonable. Figure 4 shows the estimate as a function of the interval choice. It is clear that the estimate is relatively insensitive to the exact interval choice, as long as it is reasonable.

*Estimation of $d_1$ and $d_2$.* Using functions of the CDF, we obtain expressions for $d_1$ and $d_2$. Let $t_1, t_2, t'_2$ and $t_3$ be such that $0 \le t_1 \le d_1$, $d_1 \le t'_2 < t_2 \le T - d_2$, and $T - d_2 \le t_3 < T$. For $d_1$, we use the ratio $\frac{F_3(t_2) - F_3(t'_2)}{F_3(t_1)}$ and for $d_2$, we use the ratio $\frac{F_3(t_2) - F_3(t'_2)}{1 - F_3(t_3)}$. These lead to the following expressions:

$$(11) \quad d_1 = T - T \left\{ \frac{\alpha_1}{\alpha_2} \cdot \frac{F(t_1)}{F(t_2) - F(t'_2)} \cdot \frac{(1 - t'_2/T)^{\alpha_2} - (1 - t_2/T)^{\alpha_2}}{1 - (1 - t_1/T)^{\alpha_1}} \right\}^{1/(\alpha_2 - \alpha_1)},$$

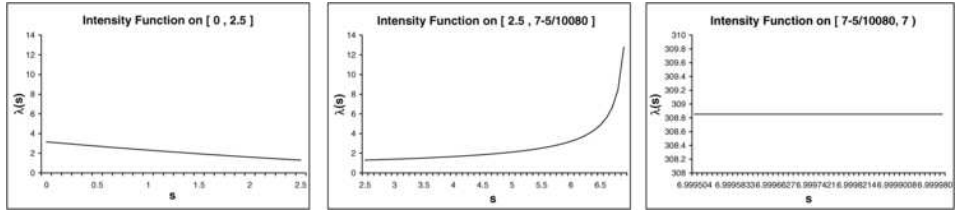

FIG. 2. *Intensity function $\lambda(s)$ for simulated data, where $\alpha_1 = 3$, $\alpha_2 = 0.4$, $\alpha_3 = 1$, $d_1 = 2.5$, $d_2 = 7 - 5/10080$, and $c = 1$. Note the different time scale for the last 5 minutes (right panel).*

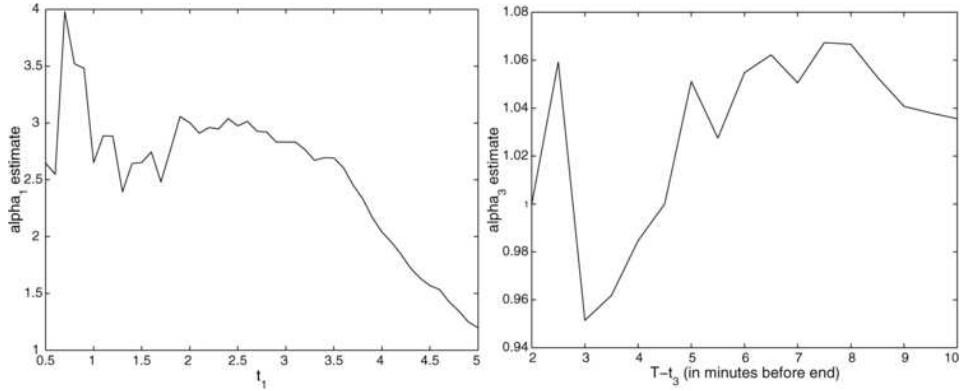

FIG. 3. *Quick estimates of $\alpha_1, \alpha_2$ and $\alpha_3$ as a function of the input intervals, for simulated BARISTA process data.*

$$(12) \quad d_2 = T \left\{ \frac{\alpha_3}{\alpha_2} \cdot \frac{1 - F(t)}{F(t_2) - F(t'_2)} \cdot \frac{(1 - t'_2/T)^{\alpha_2} - (1 - t_2/T)^{\alpha_2}}{(1 - t_3/T)^{\alpha_3}} \right\}^{1/(\alpha_2 - \alpha_3)}.$$

We can therefore estimate $d_1$ and $d_2$ by selecting "safe" values for $t_1, t'_2, t_2$ and $t_3$ (which are confidently within the relevant interval) and using the empirical CDF at those points.

Using this method, we estimated $d_1$ and $d_2$ for the simulated data. We used the true values of the $\alpha$ parameters and the "safe" values $t_1 = 1, t'_2 = 3, t_2 = 6$, and $t_3 = 7 - 2/10080$. The estimates and their (bootstrap) standard errors are reported in Table 1. Figure 5 shows the robustness of the estimates to the choice of the "safe" values. It can be seen that $d_1$ estimates are stable between 2.4–2.6 even if we choose $t_1$ slightly outside of the first interval $[0, 2.5]$. $d_2$ estimates are between 3–5.5 minutes even when $t_3$ is dislocated by a few minutes into the second interval.

*Estimation of c.* The estimate of the parameter $c$ is based on the estimate $\widehat{\theta}$ of the other parameters $\theta = (\alpha_1, \alpha_2, \alpha_3, d_1, d_2)$, and the observed number $N(T)$ of bids placed on $[0, T]$. Define $g(\theta; s) = \lambda(s)/c$, $0 \leq s \leq T$, where $\lambda$ is the function in (1). We have

$$N(T) \approx E[N(T)] = c \int_0^T g(\theta; s) \, ds \approx c \int_0^T g(\widehat{\theta}; s) \, ds.$$

Solving for $c$, we obtain the estimate

$$\widehat{c} = \frac{N(T)}{\int_0^T g(\widehat{\theta}; s) \, ds}.$$

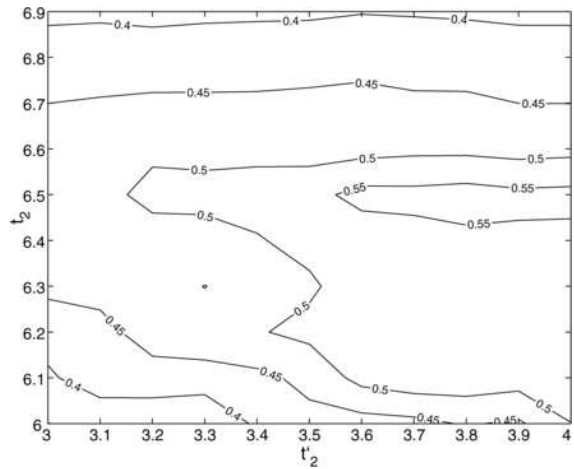

FIG. 4. *Quick and crude estimate of $\alpha_2$ as a function of $[t'_2, t_2]$ choice. $\widehat{\alpha}_2$ is between 0.4–0.55 in the entire range of intervals. The more extreme intervals ($t'_2 < 3.4$ or $t_2 > 6.8$) yield $\widehat{\alpha}_2 = 0.4$.*

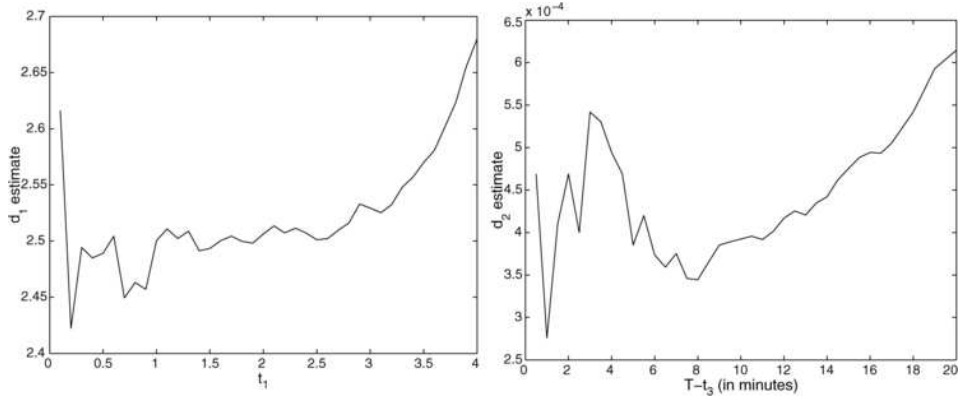

FIG. 5. *Graphs of $\hat{d}_1$ vs. $t_1$ (left) and $\hat{d}_2$ vs. initial values of $T - t_3$ (right) for simulated data. The estimate for $d_1$ is stable at $\approx 2.5$. $\hat{d}_2$ using the last 2–5 minute interval is in the range of 4–5 minutes.*

If $\hat{\theta}$ is an MLE of $\theta$, then $\hat{c}$ is an MLE of $c$.

4.2.2. *Maximum likelihood estimation.* Conditional on $N(T) = n$ (see Appendix A for unconditional estimation), the BARISTA likelihood function is given by

$$
\begin{aligned}
&\mathcal{L}(x_1, \ldots, x_n | \alpha_1, \alpha_2, \alpha_3, d_1, d_2) \\
(13)\quad &= n \ln C + n_1(\alpha_2 - \alpha_1) \ln\left(1 - \frac{d_1}{T}\right) + n_3(\alpha_2 - \alpha_3) \ln \frac{d_2}{T} \\
&\quad + (\alpha_1 - 1)S_1 + (\alpha_2 - 1)S_2 + (\alpha_3 - 1)S_3,
\end{aligned}
$$

where $n_1$ is the number of arrivals before time $d_1$, $n_3$ is the number of arrivals after $T - d_2$, $S_1 = \sum_{i:x_i \leq d_1} \ln(1 - \frac{x_i}{T})$, $S_2 = \sum_{i:d_1 < x_i < T - d_2} \ln(1 - \frac{x_i}{T})$ and $S_3 = \sum_{i:x_i > T - d_2} \ln(1 - \frac{x_i}{T})$.

In order to estimate $\alpha_1, \alpha_2, \alpha_3$ for given values of $d_1, d_2$, the following three equations must be solved (equating the first derivatives in $\alpha_1, \alpha_2, \alpha_3$ to zero):

$$(14) \qquad S_1 = n_1 \ln\left(1 - \frac{d_1}{T}\right) - \frac{n}{C}\frac{\partial C}{\partial \alpha_1},$$

$$(15) \qquad S_2 = -n_1 \ln\left(1 - \frac{d_1}{T}\right) - n_3 \ln \frac{d_2}{T} - \frac{n}{C}\frac{\partial C}{\partial \alpha_2},$$

$$(16) \qquad S_3 = n_3 \ln \frac{d_2}{T} - \frac{n}{C}\frac{\partial C}{\partial \alpha_3},$$

where

$$(17) \quad \frac{\partial C}{\partial \alpha_1} = \frac{C^2 T}{\alpha_1^2}\left(1 - \frac{d_1}{T}\right)^{\alpha_2}\left[\left(1 - \frac{d_1}{T}\right)^{-\alpha_1}\left(1 + \alpha_1 \ln\left(1 - \frac{d_1}{T}\right)\right) - 1\right],$$

$$\frac{\partial C}{\partial \alpha_2} = \frac{C^2 T}{\alpha_1 \alpha_3 \alpha_2^2} \bigg\{ \alpha_3 \bigg(1 - \frac{d_1}{T}\bigg)^{\alpha_2}$$
$$\times \bigg[\alpha_2 \ln\bigg(1 - \frac{d_1}{T}\bigg)\bigg(\alpha_2 - \alpha_1 + \alpha_2\bigg(1 - \frac{d_1}{T}\bigg)^{-\alpha_1}\bigg) - \alpha_1\bigg]$$
$$(18) \qquad\qquad + \alpha_1\bigg(\frac{d_2}{T}\bigg)^{\alpha_2}\bigg[\alpha_3 + \alpha_2 \ln\frac{d_2}{T}(\alpha_2 - \alpha_3)\bigg]\bigg\}$$
$$= \frac{C^2 T}{\alpha_2^2}\bigg[\bigg(\frac{d_2}{T}\bigg)^{\alpha_2} - \bigg(1 - \frac{d_1}{T}\bigg)^{\alpha_2}$$
$$- \frac{\alpha_2^2}{\alpha_1}\bigg(1 - \frac{d_1}{T}\bigg)^{\alpha_2} \ln\bigg(1 - \frac{d_1}{T}\bigg)\bigg(1 - \bigg(1 - \frac{d_1}{T}\bigg)^{-\alpha_1}\bigg)$$
$$- \alpha_2\bigg(\frac{d_2}{T}\bigg)^{\alpha_2} \ln\frac{d_2}{T}$$
$$+ \alpha_2\bigg(1 - \frac{d_1}{T}\bigg)^{\alpha_2} \ln\bigg(1 - \frac{d_1}{T}\bigg) + \frac{\alpha_2^2}{\alpha_3}\bigg(\frac{d_2}{T}\bigg)^{\alpha_2} \ln\frac{d_2}{T}\bigg],$$
$$(19) \quad \frac{\partial C}{\partial \alpha_3} = \frac{C^2 T}{\alpha_3^2}\bigg(\frac{d_2}{T}\bigg)^{\alpha_2}.$$

Since the equations are nonlinear in the parameters, an iterative gradient method can be used (the second derivatives are given in Appendix B). This can be solved using an iterative gradient-based method such as Newton Raphson or the Broyden–Fletcher–Goldfarb–Powell (BFGP) method, which is a more stable quasi-Newton method that does not require the computation and inversion of the Hessian matrix [see, e.g., Dennis and Schnabel (1983)]. If the changepoints $d_1$ and $d_2$ are unknown and we want to estimate them from the data, then search algorithms such as genetic algorithms can be more efficient, more stable and more easily programmable for finding a solution. Otherwise, the likelihood needs to be computed for a grid of $d_1 \times d_2$ values. In addition, empirical evidence suggests that gradient methods tend to be unstable for solving this maximization problem. In short, an exhaustive search over a reasonable grid of the parameter space or a stochastic search algorithm are good practical solutions. A good starting value would be the estimate obtained from the quick and crude method.

*Genetic algorithm search.* An alternative to an exhaustive search is the genetic algorithm (GA). The genetic algorithm belongs to a general class of stochastic, global optimization procedures that imitate the evolutionary process of nature. The basic building blocks of GA are *crossover*, *mutation* and *selection*—similar to their biological counterparts found in the evolution of genes. GA is an iterative process and each iteration is called a new generation. Starting from a parent population, two parents create offspring via crossover and mutation. The crossover operator imitates the mixing of genetic information during sexual reproduction. The mutation operator imitates the occasional changes of genetic information due to external influences. An offspring's fitness is evaluated relative to an objective function. The offspring with the highest fitness is then selected for further reproduction. Although GAs' operations appear heuristic, Holland (1975) provides theoretical arguments for convergence to a high quality optimum.

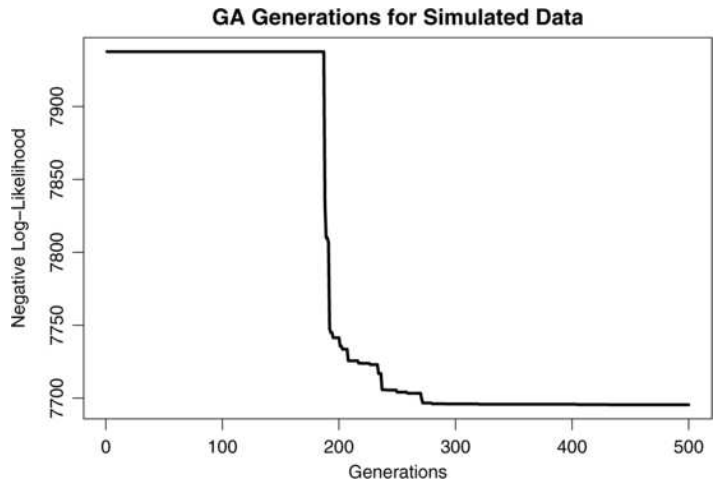

Fig. 6. *500 generations of the GA for the simulated data.*

We use a GA to estimate the parameters of the BARISTA process as follows. After creating the parent population of size $S = 100$, we select the top 10% of parents with the highest fitness and perform crossover and mutation on a randomly chosen pair, thereby creating a pair of new offspring. We repeat this 50 times to obtain an offspring population of the same size $S$ as the parent population. After creating a set of suitable offspring, the next step is to evaluate an offspring's fitness. One approach is to evaluate an offspring's fitness according to its likelihood value. Let $\theta$ denote an offspring and let $L(\theta) = L(x_1, \ldots, x_n | \theta)$ denote the corresponding likelihood value. For two offspring $\theta_1$ and $\theta_2$, $\theta_1$ has higher fitness if $\mathcal{L}(\theta_1) > \mathcal{L}(\theta_2)$.

We ran a GA on the simulated data, restricting the range of possible solutions to the hypercube $(\alpha_1, \alpha_2, \alpha_3, d_1, d_2) \in [1, 15] \times [0.1, 1] \times [0.5, 15] \times [1, 5] \times [0, 0.01]$, and running it for a total of 500 generations. Figure 6 shows a graph of the fitness-history over the 500 generations. We see that after generation 300, there are barely any further improvements. Our parameter estimates are taken from the last generation. This yielded the estimates and standard errors given in Table 1. All of these estimates are in line with the quick and crude estimates, and are very close to the values used to generate the data. The combined numerical maximization and grid search procedure did not converge.

4.2.3. *Model selection.* Although we posit that a three-stage model is, at least in general, most suitable for describing the bid arrival process in online auctions, it is possible to extend the estimation process to include model selection. To allow for a more flexible family of distributions, we consider the family of one-stage (NHPP$_1$), two-stage (NHPP$_2$) and 3-stage (BARISTA) models. Since the first two are nested within the BARISTA model, we can choose the best model using likelihood-ratios. To compare a 3-stage with a 2-stage model, for instance, we use the statistic

$$(20) \qquad -2\{\mathcal{L}(NHPP_2) - \mathcal{L}(BARISTA)\},$$

where $\mathcal{L}(i)$ is the log-likelihood for model $i$. Under the null hypothesis that the models are equivalent in their ability to fit the data (i.e., the NHPP$_2$ is sufficient), the statistic follows a $\chi^2(p)$ distribution with $p = 5 - 3 = 2$ degrees of freedom (the difference in the number of parameters of the two models). If the $p$-value is sufficiently small, then it is reasonable to choose the 3-stage model, whereas a large $p$-value would indicate the use of the 2-stage model. A similar statistic can be designed to test the difference between the 1-stage and 2-stage models, which would also follow a $\chi^2$ distribution, again with $p = 3 - 1 = 2$ degrees of freedom.

This test statistic can be used in conjunction with any of the estimation methods that we described. The most comprehensive and computationally intensive option is to find the "best" 1-stage, "best" 2-stage and "best" 3-stage models (in the sense of the highest likelihood values), and compare them using the likelihood-ratio test. A more practical alternative is to combine the model selection with a stochastic search algorithm. In particular, we incorporated model selection into the genetic algorithm as follows. Starting with the simplest model, $\text{NHPP}_1$, we apply the GA to obtain the parameter estimates and the associated log-likelihood value, $\mathcal{L}_1$. $\text{NHPP}_1$ is the most parsimonious model and we only move to a more complex model if the data justify that choice. We hence continue by fitting an $\text{NHPP}_2$ and obtaining the log-likelihood value $\mathcal{L}_2$. We compute the likelihood-ratio statistic (20) and the associated $p$-value. If the statistic is significant ($p$-value $< 0.05$), then we discard the current model ($\text{NHPP}_1$), move to the better model ($\text{NHPP}_2$), and repeat the process by comparing that model with the next model (BARISTA). Alternatively, if the likelihood-ratio statistic is insignificant ($p$-value $> 0.05$), we stop and retain the current model as the best model.

Table 1
*True and estimated values (with standard errors) for the BARISTA model parameters, by method*

|  | $\hat{\alpha}_1$ | $\hat{\alpha}_2$ | $\hat{\alpha}_3$ | $\hat{d}_1$ | $\hat{d}_2$ (minutes) |
| --- | --- | --- | --- | --- | --- |
| Simulated (true) values | 3 | 0.4 | 1 | 2.5 | 5 |
| CDF-based Q&C | 2.85 (0.06) | 0.443 (0.001) | 0.954 (0.0132) | 2.5 (0.0036) | 4.7 (0.13) |
| Genetic algorithm | 2.88 (0.007) | 0.387 (0.005) | 0.997 (0.009) | 2.63 (0.0044) | 4.6 (0.42) |

Table 2
*Information for 9 datasets on three different types of items and three different auction durations*

|         |           | 7-day | 5-day | 3-day |
|---------|-----------|-------|-------|-------|
| Xbox    | #bids     | 1861  | 393   | 557   |
|         | #auctions | 93    | 21    | 35    |
| Palm    | #bids     | 3832  | 869   | 1216  |
|         | #auctions | 194   | 54    | 35    |
| Cartier | #bids     | 1348  | 355   | 250   |
|         | #auctions | 97    | 21    | 18    |

## 5. Empirical results.

5.1. *Data.* We collected data from eBay.com on closed auctions for three types of products: Palm M515 personal digital assistants, Microsoft Xbox games and Cartier premium wristwatches. The data include auctions of three different durations: 3-day, 5-day and 7-day auctions. Relevant statistics are given in Table 2.

5.2. *Estimation.* We describe the estimation process only for the 7-day Palm bid arrival times. The same approach was used to estimate parameters for all other datasets, and we report the estimate for the entire dataset in the end.

5.2.1. *Initial quick and crude estimation.* Based on previous empirical results, we chose the first day for estimating $\alpha_1$, that is, we believe that bids placed during the first day are contained within the first "early bidding" stage. Looking at the estimate as a function of the interval chosen (Figure 7, left panel), we see that the estimate is between 4–5 if we use the first 1–2 days. It is interesting to note that after the first two days, the estimate decreases progressively reaching $\hat{\alpha}_1 = 2.5$ on the interval $[0.01, 3]$, indicating that the changepoint $d_1$ is around 2.

The parameter $\alpha_3$ was estimated using (10) with $t'_3 = 7 - 0.1/10080$ and a range of values for $t_3$. From these, $\alpha_3$ appears to be approximately 1. It can be seen in the right panel of Figure 7 that this estimate is relatively stable within the last 10 minutes. Also, notice that selecting $t_3$ too close to $t'_3$ results in unreliable estimates (due to a small number of observations between the two values).

Finally, we chose the interval $[3, 6.9]$ for estimating $\alpha_2$. This yielded the estimate $\hat{\alpha}_2 = 0.36$. Figure 8 shows the estimate as a function of the interval choice. Note that the estimate is stable between 0.2–0.4 for the different intervals chosen. It is more sensitive to the choice of $t_2$, the upper bound of the interval, and thus an overly conservative interval could yield large inaccuracies.

Using these estimates ($\hat{\alpha}_1 = 4.3, \hat{\alpha}_2 = 0.36, \hat{\alpha}_3 = 1$), we estimated $d_1$ and $d_2$. Figure 9 shows graphs of the estimates as a function of the intervals selected. The estimate for $d_1$ (left panel) appears to be stable at approximately $\hat{d}_1 = 1.75$. The estimate for $d_2$ (right panel) appears to be around 2 minutes. From the increasing values obtained for $T - t_3 > 3$ minutes, we also learn that $d_2 < 3$.

5.3. *Further refinement: ML and GA.* Table 3 displays the above estimates and compares them to the two other estimation methods: An exhaustive search over a reasonable range of the parameter space (around the quick and crude estimates) and the much quicker genetic algorithm. We restricted the range of possible solutions for the genetic algorithm to the hypercube $(\alpha_1, \alpha_2, \alpha_3, d_1, d_2) \in [0, 10] \times [0, 1] \times [0, 5] \times [0, 5] \times [0, 1000 \text{ min}]$. It can be seen that all methods yielded estimates in the same vicinity.

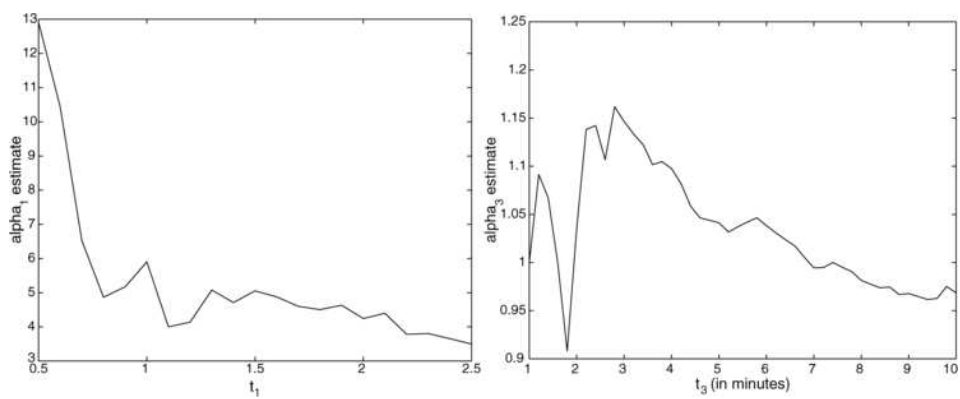

FIG. 7. *Quick and crude estimates of $\alpha_1$ as a function of $t_1$ (with $t'_1 = 0.001$) (left) and of $\alpha_3$ as a function of $t_3$ (with $t'_3 = 0.5/10080$) (right). $\hat{\alpha}_1$ is stable around 5 for $t_1$ in the range 0.75–1.75 days. A shorter interval does not contain enough data. A longer interval leads to a drop in the estimate, indicating that $d_1 < 2$. $\hat{\alpha}_3$ is around 1.1 when $t_3$ is within the last 2–4 minutes.*

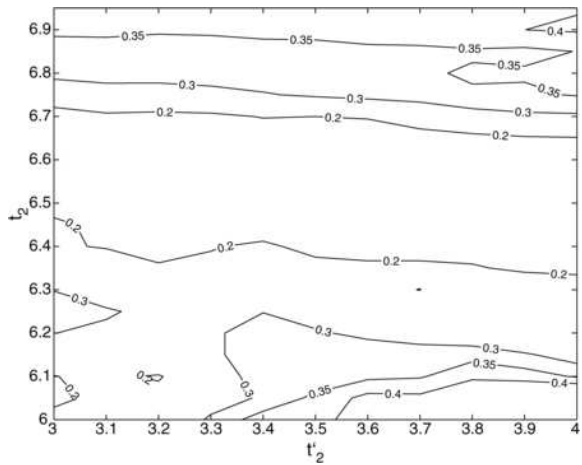

FIG. 8. *Quick and crude estimate of $\alpha_2$ as a function of $[t'_2, t_2]$. Shorter, "safer" intervals are at the lower right. Longer intervals, containing more data, are at the upper left. $\hat{\alpha}_2$ is between 0.2–0.4 for all intervals. For $t_2 > 6.9$, the estimate is approximately 0.35.*

We also performed model selection to see whether a 2-stage or 1-stage model would sufficiently fit the data. The low $p$-values for comparing the 3-stage model with these models showed that indeed a 3-stage model is preferable and more accurately approximates the data. For a detailed fitting of the data to an $\text{NHPP}_1$ (which includes the Poisson as a special case) and an $\text{NHPP}_2$, see Shmueli et al. (2004).

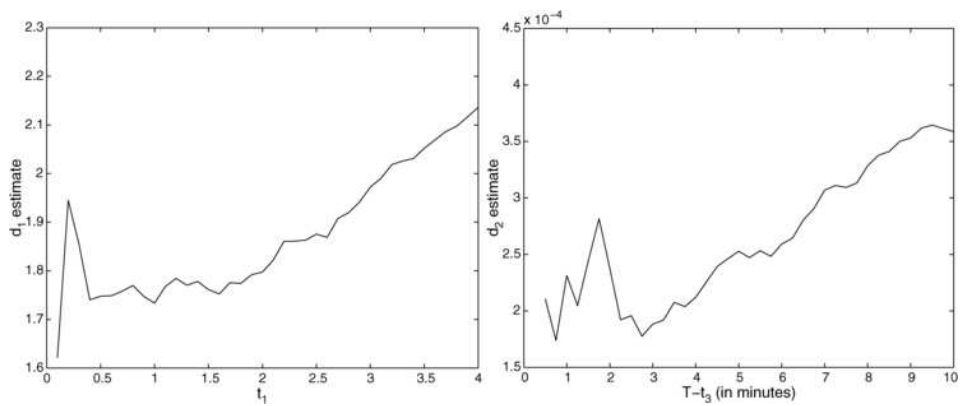

FIG. 9. *Plots of $\hat{d}_1$ vs. $t_1$ (left) and $\hat{d}_2$ vs. initial values of $T - t_3$ (right) for Palm data. The estimate for $d_1$ seems stable at $\approx 1.75$. $\hat{d}_2$ is approximately 2 minutes.*

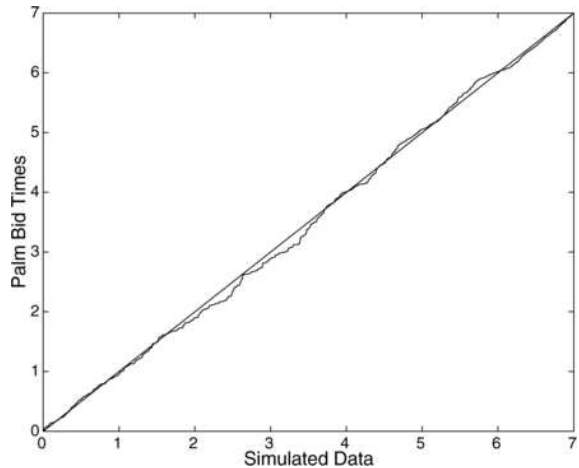

Fig. 10. *Q–Q plot of Palm bid times vs. simulated data from a BARISTA process with parameters* $\alpha_1 = 4.9, \alpha_2 = 0.37, \alpha_3 = 1.13, d_1 = 1.7, d_2 = 2/10080$.

Finally, to further validate this estimated model, we simulated data from a BARISTA process with the above ML estimates as parameters, and number of bids equal to that in the respective dataset. Figure 10 shows a QQ-plot of the Palm data vs. the simulated data. The points appear to fall on the line $x = y$, thus, supporting the adequacy of the estimated model for the Palm bid times.

The estimated model for the Palm data reveals the dynamics of these auctions over time: Indeed, the "average" auction has three stages: the initial stage takes place during the first 1.7 days, the middle stage continues until the last 2 minutes, and then the third stage kicks in. The bid arrivals in each of the three stages have different intensity functions. The auction beginning is characterized by an early surge of interest, with more intense bidding than during the start of the second stage. Then, the increase in bid arrival rate slows down during the middle of the auction. The bids do tend to arrive faster as the auction progresses, but at the very end, during the last 2 minutes of the auction, we observe a uniform bid arrival process. Finally, it is interesting to note that in these data the third stage of bidding seems to take place within the last 2–3 minutes compared to the last 1 minute in Roth and Ockenfels (2000). Thus, we use the term "last-moment bidding" rather than "last-minute bidding."

TABLE 3
*Estimates for five BARISTA model parameters using the three estimation methods*

|  | $\hat{\alpha}_1$ | $\hat{\alpha}_2$ | $\hat{\alpha}_3$ | $\hat{d}_1$ | $\hat{d}_2$ (minutes) |
|---|---|---|---|---|---|
| CDF-based Q&C | 4.3 (0.02) | 0.36 (0.001) | 1 (0.02) | 1.8 (0.009) | 3.3 (0.18) |
| Exhaustive search | 4.9 | 0.37 | 1.13 | 1.7 | 2.0 |
| Genetic algorithm | 5.55 (0.005) | 0.35 (0.005) | 1.1 (0.01) | 1.55 (0.005) | 2.0 (0.10) |

5.4. *Comparing auctions.* Table 4 gives the estimated BARISTA coefficients for all nine datasets. In all cases the fit was obtained by the process described above, using a GA-based model selection procedure and refining the estimates using an exhaustive search over the likelihood.

We can see that except for one dataset (5-day Xbox auctions), a three-stage (BARISTA) model provided the best fit among the one-, two- and three-stage models. Also, in all cases the final fit of the model to the data, evaluated by examining QQ-plots, was very good. The last phase lasted a couple of minutes or less in all auctions, irrespective of the auction duration, and there was more consistency within a certain item than within a certain duration. The first stage tends to last 1–2 days, with one exception being 4 days (7-day Cartier auctions). Another common feature is the magnitude of $\alpha_2$, around 0.3, indicating that during the second phase the bidding frequency is equivalently low in different items' auctions and in different durations. The bidding intensity during the last short stage, $\alpha_3$, is also typically around 1, with two datasets reaching nearly 3.

The parameter that varies most across datasets is $\alpha_1$, the "early bidding" frequency. It appears that the 3-day auctions exhibit a lower level of early bidding compared to the longer duration auctions. However, there still does exist such a stage even in these "short" auctions.

Finally, from a computational complexity point, the GA with model-selection ran reasonably fast, with the longest estimation taking 11 minutes. The runtimes are summarized in Table 5.

Table 4
*Estimated BARISTA coefficients for all 9 datasets, using a GA-based model selection and an exhaustive search over the likelihood function of the selected set of models*

|  |  | # stages | $\hat{\alpha}_1$ | $\hat{\alpha}_2$ | $\hat{\alpha}_3$ | $\hat{d}_1$ | $\hat{d}_2$ (min) |
|---|---|---|---|---|---|---|---|
|  | 7-days | 3 | 10 | 0.24 | 1.19 | 1.2 | 2 |
| Xbox | 5-days | 2 | – | 0.3 | 7.7 | – | 1 |
|  | 3-days | 3 | 2.1 | 0.3 | 2.99 | 1.3 | 1.8 |
|  | 7-days | 3 | 4.9 | 0.37 | 1.13 | 1.7 | 2 |
| Palm | 5-days | 3 | 2.7 | 0.32 | 0.79 | 1.8 | 2 |
|  | 3-days | 3 | 1.8 | 0.33 | 1.01 | 1.45 | 1.4 |
|  | 7-days | 3 | 2.6 | 0.29 | 2.95 | 4 | 0.5 |
| Cartier | 5-days | 3 | 3.1 | 0.35 | 1.08 | 2.4 | 0.51 |
|  | 3-days | 3 | 1.5 | 0.31 | 0.88 | 1.8 | 1 |

TABLE 5
*Computational complexity of the GA: The table lists the run-time of the GA (recorded in minutes) for the 9 data sets described in Section 5*

| Data  | Palm  | XBox | Cartier |
|-------|-------|------|---------|
| 3-day | 3.27  | 1.35 | 0.45    |
| 5-day | 2.19  | 1.09 | 1.03    |
| 7-day | 11.27 | 5.34 | 3.54    |

**6. Relating bidder arrivals and bid arrivals.** The online auction literature is rich with papers that assume an ordinary homogenous Poisson bidder arrival process. This assumption underlies various theoretical derivations, is the basis for the simulation of bid data, and is used to design field experiments. Bajari and Hortacsu (2000) specify and estimate a structural econometric model of bidding on eBay, assuming a Poisson bidder arrival process. Etzion et al. (2003) suggest a model for segmenting consumers at dual channel online merchants. Based on the assumption of Poisson arrivals to the website, they model consumer choice of channel, simulate consumer arrivals and actions, and compute relationships between auction duration, lot size and the constant Poisson arrival rate $\lambda$. Zhang et al. (2002) model the demand curve for consumer products in online auctions based on Poisson bidder arrivals, and fit the model to bid data. Pinker et al. (2003) and Vakrat and Seidmann (2000) use a Poisson process for modeling the arrival of bidders in going-going-gone auctions. They use the intensity function $\lambda(t) = \lambda_a e^{-t/T}$, $0 \leq t \leq T$, where $T$ is the auction duration, and $\lambda_a$ is the intensity of website traffic into the auction. This model describes the decline in the number of new bidders as the auction progresses. Haubl and Popkowski Leszczyc (2003) design and carry out an experiment for studying the effect of fixed-price charges (e.g., shipping costs) and reserve prices on consumer's product valuation. The experiment uses simulated data that are based on Poisson arrivals of bidders. These studies are among the many that rely on a Poisson arrival process assumption.

In online auctions, however, bidder arrivals are unobserved as we pointed out earlier. Therefore, it is not straightforward to study their distribution. On the other hand, bid arrivals *are* observed. In the following we investigate the relationship between bidder arrivals and bid arrivals more carefully.

6.1. *Poisson bidder arrivals yield NHPP bid arrivals.* We now establish a key connection between the bidder arrival and bid arrival processes. Suppose that bidders enter an auction in accordance with a Poisson process having a fixed rate $\lambda$, and that a bidder who arrives at time $s$ places a single bid on the interval $[s, T)$ according to a bid time distribution $G_s$. The resulting bid arrival process is similar to the output process of an $M/G/\infty$ queue, except that the *service time* (the elapsed time between a bidder's arrival and the placement of his bid) is dependent on the arrival time of the bidder.

By Proposition 2.3.2 of Ross (1995), the bid counts on nonoverlapping subintervals of $[0, T]$, are independent variables, and hence the bid arrival process possesses independent increments. Moreover, the bid count on $[0, y]$ is *Poisson* distributed with mean

$$\int_0^y \lambda G_s(y)\,ds = \int_0^y \lambda \left[\int_s^y dG_s(t)\,dt\right] ds$$
$$= \int_0^y \left[\lambda \int_0^t dG_s(t)\,ds\right] dt$$
$$= \int_0^y \left[\lambda \int_0^t g_s(t)\,ds\right] dt \qquad \text{(if } G_s \text{ has a derivative } g_s\text{)},$$

and hence, if $G_s$ has derivative $g_s$, then the bid arrival process is a nonhomogeneous Poisson process with intensity function $\lambda(t) = \lambda \int_0^t g_s(t)\,ds$. The function $g_s$ is the link between the bidder arrival process and the resulting bid arrival process. Suppose, for example, that a bidder who arrives at time $s$ places a single bid uniformly on the remaining interval $(s, T)$. Then, $g_s(t) = 1/(T-s)$ so that $\lambda(t) = \lambda \log(T/(T-t))$. Thus, fixed rate Poisson (uniform) bidder arrivals, in conjunction with uniform placement of their single bids, yields a nonhomogeneous Poisson process of bid arrivals with an intensity that increases as the auction deadline approaches.

6.2. *Poisson bidder arrivals yield BARISTA bid arrivals.* Continuing the discussion in Section 6.1, we describe a naturally arising bid time distribution that yields a BARISTA bid arrival process. Suppose that a bidder who arrives at time $s_1$ places his bid immediately (at time $s_1$) with probability $\alpha > 0$, and makes no further bids, or otherwise selects $s_2 \sim U(s_1, T)$ as his next *potential* bid time. At time $s_2$ he again places his bid with probability $\alpha$, and makes no further bids, or otherwise selects $s_3 \sim U(s_2, T)$ as his next *potential* bid time. Continuing in this manner, the bidder eventually places his single bid and then departs the auction. By the discussion in Section 6.1, the resulting bid arrival process is a nonhomogeneous Poisson process.

To derive the form of the intensity function, let $\tau$ denote the bid time of a randomly chosen bidder. The intensity function $\lambda$ is a constant multiple of $\frac{d}{dt}P(\tau \leq t)$. By Lemma 1 of Appendix C (set $a = 0$ and $b = T$), we have

$$\lambda(t) = (const)\frac{d}{dt}P(\tau \leq t) = (const)\frac{d}{dt}\left\{1 - \left(1 - \frac{t}{T}\right)^\alpha\right\}$$
(21)
$$= (const)\left(1 - \frac{t}{T}\right)^{\alpha-1}, \qquad 0 \leq t \leq T.$$

This is the intensity function of our $\text{NHPP}_1$.

To obtain $\text{NHPP}_2$ bid arrivals with $\alpha_2 = \alpha$ and $\alpha_3 = 1$, we alter the above set-up so that a bidder who selects a *potential* bid time after $T - d$ either places a bid at that time (with probability $\alpha$) or otherwise leaves the auction without placing a bid (or the one bid he attempts to place on $[T-d, T]$ fails to transmit). Note that this variation produces uniformly distributed bid arrivals on $(T - d, T]$.

To generalize further to $\text{NHPP}_2$ bid arrivals with $0 < \alpha_2 \leq \alpha_3 \leq 1$, we replace $\alpha$ by $\alpha_2$ on $[0, T - d]$, and by $\alpha_3$ on $(T - d, T]$. We also assign probability $1 - \alpha_2/\alpha_3$ to the event that a bidder who has not placed his bid by time $T - d$ ultimately departs the auction without bidding. Let $\tau$ denote the bid time of a bidder chosen randomly from those who successfully place a bid. By multiple applications of Lemma 1 of Appendix C,

$$P(\tau \leq t) = \begin{cases} \dfrac{1}{\pi}\left\{1 - \left(1 - \dfrac{t}{T}\right)^{\alpha_2}\right\}, & \text{for } 0 \leq t < T - d, \\ 1 - \dfrac{1}{\pi}\dfrac{\alpha_2}{\alpha_3}\left(\dfrac{d}{T}\right)^{\alpha_2}\left(\dfrac{T - t}{d}\right)^{\alpha_3}, & \text{for } T - d \leq t \leq T, \end{cases}$$

where $\pi$ denotes the probability that a randomly chosen bidder successfully places a bid [$\pi$ is equal to $1 - (\frac{d}{T})^{\alpha_2}(1 - \frac{\alpha_2}{\alpha_3})$]. We have for some $c > 0$

$$\lambda(t) = (const)\dfrac{d}{dt}P(\tau \leq t)$$

$$= \begin{cases} c\left(1 - \dfrac{t}{T}\right)^{\alpha_2 - 1}, & \text{for } 0 \leq t < T - d, \\ c\left(\dfrac{d}{T}\right)^{\alpha_2 - \alpha_3}\left(1 - \dfrac{t}{T}\right)^{\alpha_3 - 1}, & \text{for } T - d \leq t \leq T. \end{cases}$$

This is the intensity function of our $\text{NHPP}_2$.

**7. Discussion.** Empirical research on bid timing in online auctions has been exploratory and data-driven. The BARISTA model is the first proposed probabilistic model for the bid arrival process in online auctions. This probabilistic foundation provides an improved platform for quantifying bid arrival processes and for simulating data, and is a first step for establishing models of bidder strategies (as shown in Section 6). It can also be used to improve nonparametric representations of price-processes in online auctions [e.g., used for clustering auctions in Jank and Shmueli (2007) or for forecasting ongoing auctions in Wang et al. (2007)], by specifying the amount and locations of knots in the smoothing splines.

One possible extension of the current BARISTA formulation is to auctions of random duration. An alternative to the popular fixed-length format (as on eBay) is a format whereby the closing time is extended, beyond the scheduled deadline if necessary, until a predetermined number of minutes have passed without a bid being submitted. This is known as a popcorn ending (as on Amazon.com). Extending the BARISTA model in this direction is likely to require an additional fourth phase and a modification to the bid intensity in the third phase, as the parameter $T$ has a different role in the alternative format. Formulating such a model and fitting it to real data is an interesting future direction. An enhanced BARISTA that models the bivariate process of bid arrivals and bid amounts is also of interest, as is the probabilistic formulation of bidder arrivals and strategies that lead to BARISTA bid arrivals.

## APPENDIX A: ML ESTIMATION OF THE UNCONDITIONAL BARISTA MODEL

Let $N(s), 0 \leq s \leq T$, be a NHPP with an intensity function of the form

$$\lambda(s) = cg(\theta, s), \qquad 0 \leq s \leq T,$$

where $c$ and $\theta = (\theta_1, \ldots, \theta_k)$ are unknown parameters. Define $h(\theta) = \int_0^T g(\theta, s)\, ds$, so that $m(T) = ch(\theta)$. The *pdf* associated with $\lambda$ is $f(\theta, s) = \lambda(s)/m(T)$, $0 \leq s \leq T$. Given a random sample $x_1, \ldots, x_n$ (nonrandom $n$) from this distribution, the likelihood and log-likelihood functions of $\theta$ are

$$L(\theta) = \prod_{i=1}^{n} f(\theta, x_i) \quad \text{and} \quad \mathcal{L}(\theta) = \ln L(\theta).$$

On the other hand, given the value $n$ of $N(T)$, and the arrival times $x_1, \ldots, x_n$ from the NHPP, the likelihood function of $(c, \theta)$ is given by

$$L(c, \theta) = \frac{e^{-m(T)} m(T)^n}{n!} \prod_{i=1}^{n} f(\theta, x_i) = \frac{e^{-ch(\theta)}(ch(\theta))^n}{n!} L(\theta).$$

The log-likelihood is thus

$$\mathcal{L}(c, \theta) = -ch(\theta) + n \ln c + n \ln h(\theta) - \ln n! + \mathcal{L}(\theta).$$

The joint MLE of $c$ and $\theta$ is the solution of the equations

(A.1)
$$0 = \frac{\partial \mathcal{L}(c, \theta)}{\partial c} = -h(\theta) + \frac{n}{c},$$
$$0 = \frac{\partial \mathcal{L}(c, \theta)}{\partial \theta_j} = -c \frac{\partial h(\theta)}{\partial \theta_j} + \frac{n}{h(\theta)} \frac{\partial h(\theta)}{\partial \theta_j} + \frac{\partial \mathcal{L}(\theta)}{\partial \theta_j}, \qquad 1 \leq j \leq k.$$

Solving the first equation in (A.1) for $c$ and plugging into the second, we find that

$$\frac{\partial \mathcal{L}(c, \theta)}{\partial \theta_j} = \frac{\partial \mathcal{L}(\theta)}{\partial \theta_j}, \qquad 1 \leq j \leq k.$$

Hence, $L(c, \theta)$ and $L(\theta)$ yield the same MLE for $\theta$. That is, if $\widehat{\theta}_j = w_j(X_1, \ldots, X_n)$ is the *MLE* of $\theta_j$ $(1 \leq j \leq k)$ based on a random sample of nonrandom size $n$ from the distribution with the *pdf* above, then the *MLE* of $\theta_j$ based on the arrival times $X_1, \ldots, X_{N(T)}$ from the above NHPP is of the form $\widehat{\theta}_j = w_j(X_1, \ldots, X_{N(T)})$. By the first equation in (A.1), the MLE of $c$ is

$$\text{(A.2)} \qquad \widehat{c} = \frac{N(T)}{h(\widehat{\theta})}.$$

## APPENDIX B: SECOND DERIVATIVES OF THE LOG-LIKELIHOOD FUNCTION

The second derivatives are given for using gradient methods of ML estimation such as Newton Raphson:

$$\text{(B.1)} \quad \begin{aligned} \frac{\partial^2 \mathcal{L}}{\partial^2 \alpha_1} &= -\frac{n}{C^2}\left(\frac{\partial C}{\partial \alpha_1}\right)^2 + \frac{n}{C}\frac{\partial^2 C}{\partial^2 \alpha_1} \\ &= \frac{n}{C^2}\left(\frac{\partial C}{\partial \alpha_1}\right)^2 - \frac{n}{C}\left(\frac{2}{\alpha_1} + \ln\left(1 - \frac{d_1}{T}\right)\right)\frac{\partial C}{\partial \alpha_1}, \end{aligned}$$

$$\text{(B.2)} \quad \begin{aligned} \frac{\partial^2 \mathcal{L}}{\partial^2 \alpha_2} &= -\frac{n}{C^2}\left(\frac{\partial C}{\partial \alpha_2}\right)^2 + \frac{n}{C}\frac{\partial^2 C}{\partial^2 \alpha_2} \\ &= \frac{n}{C^2}\left(\frac{\partial C}{\partial \alpha_2}\right)^2 + \frac{2n}{\alpha_2 C}\frac{\partial C}{\partial \alpha_2} \\ &\quad - \frac{nCT}{\alpha_2}\left[\frac{1}{\alpha_3}\left(\frac{d_2}{T}\right)^{\alpha_2} \ln\frac{d_2}{T}\left(2 + (\alpha_2 - \alpha_3)\ln\frac{d_2}{T}\right)\right. \\ &\quad - \frac{1}{\alpha_1}\left(1 - \frac{d_1}{T}\right)^{\alpha_2} \ln\left(1 - \frac{d_1}{T}\right)\left(1 - \left(1 - \frac{d_1}{T}\right)^{-\alpha_1}\right) \\ &\quad \left. \times \left(2 + \alpha_2 \ln\left(1 - \frac{d_1}{T}\right)\right)\right], \end{aligned}$$

$$\text{(B.3)} \quad \frac{\partial^2 \mathcal{L}}{\partial^2 \alpha_3} = -\frac{n}{C^2}\left(\frac{\partial C}{\partial \alpha_3}\right)^2 + \frac{n}{C}\frac{\partial^2 C}{\partial^2 \alpha_3} = \frac{n}{C^2}\left(\frac{\partial C}{\partial \alpha_3}\right)^2 - \frac{n\alpha_3}{2C}\frac{\partial C}{\partial \alpha_3},$$

$$\text{(B.4)} \quad \frac{\partial^2 \mathcal{L}}{\partial \alpha_1 \alpha_2} = \frac{2}{C}\frac{\partial C}{\partial \alpha_1}\frac{\partial C}{\partial \alpha_2} + \ln\left(1 - \frac{d_1}{T}\right)\frac{\partial C}{\partial \alpha_1},$$

$$\text{(B.5)} \quad \frac{\partial^2 \mathcal{L}}{\partial \alpha_1 \alpha_3} = \frac{2}{C}\frac{\partial C}{\partial \alpha_1}\frac{\partial C}{\partial \alpha_3},$$

$$\text{(B.6)} \quad \frac{\partial^2 \mathcal{L}}{\partial \alpha_2 \alpha_3} = \frac{2}{C}\frac{\partial C}{\partial \alpha_2}\frac{\partial C}{\partial \alpha_3} + \ln\left(\frac{d_2}{T}\right)\frac{\partial C}{\partial \alpha_3}.$$

## APPENDIX C: A GEOMETRIC SERIES OF SHRINKING UNIFORM VARIABLES

LEMMA 1. *Suppose $X_1 \sim U(a,b), X_2 \sim U(X_1,b), X_3 \sim U(X_2,b), \ldots$ and $M \sim geom(\alpha)$. Then,*

$$P(X_M > s) = \left(1 - \frac{s-a}{b-a}\right)^\alpha, \qquad a \le s \le b.$$

PROOF. For convenience, assume $a = 0$ and $b = 1$. Define $p(s) = P(X_M > s)$. For small $x$,

$$p(s+x) = p(s)P(X_M > s+x | X_M > s)$$
$$\ge p(s)\left[\left(1 - \frac{x}{1-s}\right) + \left(\frac{x}{1-s}\right)(1-\alpha)\left(1 - \frac{x}{1-s}\right)\right],$$

and thus,

(C.1) $$\liminf_{x \to 0} \frac{p(s+x) - p(s)}{x} \ge \frac{-\alpha p(s)}{1-s}.$$

Moreover,

$$p(s+x) \le p(s)\left[\left(1 - \frac{x}{1-s}\right) + \frac{x(1-\alpha)}{1-s}\right],$$

so that

(C.2) $$\limsup_{x \to 0} \frac{p(s+x) - p(s)}{x} \le \frac{-\alpha p(s)}{1-s}.$$

By (C.1) and (C.2), $p'(s) = -\alpha p(s)(1-s)^{-1}$, from which we conclude that

$$p(s) = (1-s)^\alpha. \qquad \square$$

**Acknowledgments.** The authors thank Professors N. D. Shyamalkumar and R. L. Dykstra for useful discussions and the anonymous referee and AE for valuable comments. We also thank Professor Sharad Borle for the Cartier data and Shanshan Wang for the XBox data.

G. Shmueli
W. Jank
Department of Decision
  and Information Technologies
and
The Center for Electronic Markets
  and Enterprises
Robert H. Smith School of Business
University of Maryland
College Park, Maryland 20742
USA
E-mail: gshmueli@rhsmith.umd.edu
        wjank@rhsmith.umd.edu

R.P. Russo
Department of Statistics
  and Actuarial Science
University of Iowa
Iowa City, Iowa 52242
USA
E-mail: rrusso@stat.uiwoa.edu